\documentclass[12pt,prd,tightenlines,nofootinbib,showpacs,showkeys]{revtex4-2}
\usepackage{bm}
\usepackage{color,graphics}
\usepackage{rotating}
\usepackage{epsfig}
\usepackage{amssymb}
\usepackage{amsmath}
\usepackage{amsthm}
\usepackage{amsfonts}
\usepackage{amssymb}
\usepackage{braket}
\usepackage{graphics}
\usepackage{rotating}
\usepackage{indentfirst}
\usepackage[colorlinks=true,linkcolor=blue,citecolor=blue,urlcolor=blue]{hyperref}

\begin{document}

\title{Contribution of hadronic light-by-light scattering to the hyperfine structure of muonium}
\author{\firstname{V.~I.}~\surname{Korobov}}
\affiliation{BLTP JINR, Dubna, Russia}
\affiliation{Samara University, Samara, Russia}
\author{\firstname{A.~V.}~\surname{Eskin}}
\affiliation{Samara University, Samara, Russia}
\author{\firstname{A.~P.}~\surname{Martynenko}}
\affiliation{Samara University, Samara, Russia}
\author{\firstname{F.~A.}~\surname{Martynenko}}
\affiliation{Samara University, Samara, Russia}

\begin{abstract}
The contribution of hadronic scattering of light-by-light to the hyperfine structure of muonium 
is calculated using experimental data on the transition form factors of two photons into a hadron. 
The amplitudes of interaction between a muon and an electron with horizontal and vertical exchange are constructed.
The contributions due to the exchange of pseudoscalar, axial vector, scalar and tensor mesons are taken 
into account.
\end{abstract}



\maketitle

\section{Introduction}
\label{vv}

Exotic atoms such as the muonium atom, positronium atom, positronium ion, muonic hydrogen, etc. 
play a very important role in modern physics. Precise study of their energy levels, decay widths 
is such a direction of fundamental research, within which one can look for manifestations of new 
interactions of particles. Although such systems do not exist for a long time by the standards 
of conventional systems, their creation and experimental study allows one to look into a field 
of research that is inaccessible when working with stable atoms and molecules. We can say that 
the study of exotic systems, along with collider physics, is a tool for understanding reality 
beyond the Standard Model.

Electromagnetic two-particle bound states make it possible to test one of the most successful 
theories of particle interaction - quantum electrodynamics. Theoretical calculations of the energy 
levels of the simplest bound states in quantum electrodynamics have reached a very high accuracy
\cite{eides,eides1,sgk,eides1a}. 
But since the accuracy of the experimental study of energy levels has steadily increased in recent 
decades, this has led to the need to study not only the electromagnetic high order contributions 
but also the contributions of weak and strong interactions to the energy spectrum of such systems.
For example, the contribution of hadronic vacuum polarization has already reached 
the level of experimental verification for the anomalous magnetic moment (AMM) of the muon, hyperfine 
splitting in muonium, in the Lamb shift, and the hyperfine structure of muonic hydrogen.
The most acute situation with the calculation of hadronic contributions has developed for the 
AMM muon \cite{fj,radzhabov,cmd}. 
But for the other two problems, the hadronic contributions also become significant, taking into account 
the increasing precision of the experiment.

The study of the fine and hyperfine structure (HFS) of muonium has been central to the study of quantum 
electrodynamics for decades, since in this purely lepton system of different leptons there are 
no nuclear structure effects, which have always been the main theoretical uncertainty \cite{eides,eides1,sgk}. 
In recent 
years, new more accurate experimental studies related to muonium have already begun.
The Mu-MASS (MuoniuM lAser SpectroScopy)
collaboration aims to measure the
$1S-2S$ transition in muonium with a final uncertainty of 10 kHz, providing a 1000-fold improvement
on accuracy \cite{mumass}.
New result of measurement of the n=2 Lamb shift in muonium
comprises an order of magnitude improvement upon the previous best measurement  \cite{ben}.
The MuSEUM (Muonium Spectroscopy Experiment Using Microwave)
collaboration performed a new precision measurement of the muonium ground-state hyperfine 
structure at J-PARC using a high-intensity pulsed muon beam \cite{museum}.
The accuracy of the experimental result in \cite{museum} is 4 kHz and is still less than the accuracy 
of the previous experiment in 1999 \cite{lanl}.
One can consider experiments with muonium for more precise determination on the mass
ratio $m_\mu/m_e$, for the test of the Standard Model with greater accuracy and possibly,
for revealing the source of previously unaccounted interaction between particles forming the
bound state in QED. 
According to the work \cite{eides2}, the theory predicts $\nu_{HFS}=4463302872 (515)~Hz$,
$\delta=1.3\times 10^{-7}$, where the most part of the uncertainty (511 Hz) is dominated by
the measurement of the ratio $m_\mu/m_e$ (120 ppb).
Therefore, from a comparison of the theoretical and new experimental results for muonium HFS, one can obtain a more accurate 
value for the mass ratio $m_\mu/m_e$.
The MuSEUM collaboration aims to precisely measure the ground-state hyperfine splitting of muonium atoms
with the accuracy 1 ppb \cite{jparc}.

Such a high experimental accuracy of measuring the hyperfine structure of muonium at a level of 1 Hz 
requires corresponding theoretical calculations of various high-order corrections to the fine structure constant. 
Such calculations have been carried out over the years by various groups. In this paper, we study only one 
of the contributions to the hyperfine structure connected with the effect of light-by-light scattering, 
which leads to the production of various mesons in the intermediate state. 
In the quark model, such processes are determined by the production of a pair of light quarks 
and antiquarks in the $\gamma^\ast\gamma^\ast$ interaction, which can then form a light meson.
The corresponding interaction 
amplitudes are shown in Fig.~\ref{pic1}. They can be divided into two parts, which we call vertical and horizontal 
exchanges. In our previous work \cite{apm2002}, we investigated the contribution connected with the horizontal exchange 
of pseudoscalar mesons. A more complete study of these processes was carried out in Ref.~\cite{arkasha}, 
in which, along with horizontal exchanges, the contribution of vertical exchanges, including axial 
vector mesons, was also investigated. In our recent papers, we calculated the hadronic contributions 
of light-by-light scattering into the fine and hyperfine structure of muonic hydrogen \cite{apm1,apm2,apm3,apm4,apm5}
(see also Refs.~\cite{roig,frantsiska,pang,kou}) 
and showed that such processes must be taken into account when obtaining the total value of a specific 
energy interval, taking into account the ever-increasing accuracy experiments carried out 
by the CREMA collaboration \cite{crema1,crema2,crema3,crema4}, as well as other collaborations are planned
\cite{crema5,famu,famu1}. The purpose of this work is to calculate all possible meson contributions (pseudoscalar, 
scalar, axial vector and tensor) to the hyperfine splitting (HFS) in muonium and to estimate the possible 
total contribution from such interactions.
The factor determining the order of the contribution, $m_e^3\alpha^7/\Lambda^2 h\sim 0.04$~Hz where $\Lambda$ is
typical hadron mass near 1 GeV, 
is estimated to be not very large due to recoil effects and the nature 
of the hadronic interaction itself. 
Nevertheless, the study of such contributions in the hyperfine structure is of interest in connection 
with an increase in the accuracy of measurements.
Thus, for example, in the case of muonic hydrogen, hadronic effects of light-by-light scattering 
turn out to be rather significant 
both in the Lamb shift and in the hyperfine splitting \cite{apm1,apm2,apm3,apm4,apm5}.

\section{Contribution of axial vector mesons}
\label{sec2}

\begin{figure}[htbp]
\centering
\includegraphics[scale=1.]{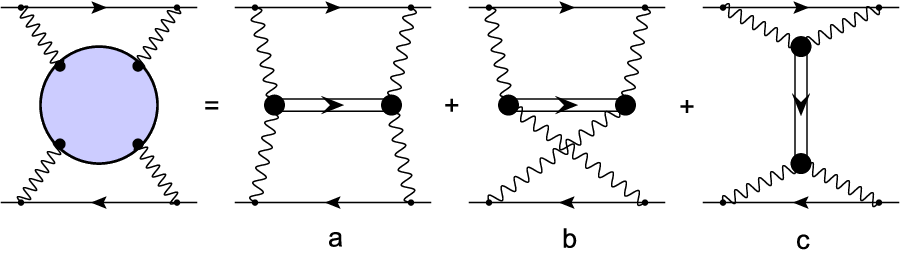}
\caption{Hadronic light-by-light scattering amplitudes with horizontal and vertical exchanges. 
Wavy line corresponds to the virtual photon.
The bold dot denotes the form factor of the transition of two photons into a meson.
}
\label{pic1}
\end{figure}

We begin the discussion of the contributions of axial vector mesons from the vertical exchange 
amplitudes in Fig.~\ref{pic1}(c).
The diagram has a vertex of the transition of two virtual photons to an axial vector meson, for which the 
following parametrization is used \cite{cahn,apm1}:
\begin{equation}
T^{\mu\nu}(k_1,k_2)=4\pi i\alpha\varepsilon_{\mu\nu\alpha\beta}(k_1^\alpha k_2^2-k_2^\alpha k_1^2)
\varepsilon_A^\beta A(t^2,k_1^2,k_2^2),
\label{f1}
\end{equation}
where $A(t^2,k_1^2,k_2^2)$ is a scalar function of the four-momentum transfer squared of the virtual 
photons $k_1^2$, $k_2^2$ describing the vertex in Fig.~\ref{pic1}.
$k_1=k$, $k_2=t-k$ are four-momenta of virtual photons, $t=p_1-q_1=(0,{\bf t})$ is the four-momentum
of the meson, $p_1$, $p_2$ are four-momenta of electron and muon in initial state,
$q_1$, $q_2$ are four-momenta of electron and muon in final state, $M_A$ is the mass of axial vector meson.
Note that the axial vector decay into two real photons is forbidden by Landau-Yang theorem
but the process with one virtual photon can already take place.
To pick out the electron-muon states with a certain spin, we use projection operators constructed from the 
wave functions of the particles in their rest frame:
\begin{equation}
\hat\Pi_{S=0}=[u(0)\bar v(0)]_{S=0}=\frac{(1+\gamma^0)}{2\sqrt{2}}\gamma_5,~~~
\hat\Pi_{S=1}=[u(0)\bar v(0)]_{S=1}=\frac{(1+\gamma^0)}{2\sqrt{2}}\hat\varepsilon.
\label{f2}
\end{equation}

As a result, the general expression for the interaction amplitude in Fig.\ref{pic1}(c) can be transformed 
to the following trace:
\begin{equation}
i{\cal M}^c=\frac{\alpha^2(Z\alpha)^2}{16m_1^2m_2^2}
\int\frac{d^4k}{\pi^2}\frac{A(t^2,k_1^2,k_2^2)}{(k^2)^2}
\int\frac{d^4r}{\pi^2}\frac{A(t^2,r_1^2,r_2^2)}{(r^2)^2}
\frac{\varepsilon_{\mu\nu\alpha\beta}(k_1^\alpha k_2^2-k_2^\alpha k_1^2)}{(k^2-2k_0m_1)}\times
\label{f3}
\end{equation}
\begin{displaymath}
\frac{\varepsilon_{\sigma\lambda\rho\omega}(r_1^\rho r_2^2-r_2^\rho r_1^2)}{(r^2-2r_0m_2)}D^{\beta\omega}(t)
Tr\Bigl[(\hat q_1+m_1)\gamma^\nu(\hat p_1-\hat k+m_1)\gamma^\mu)\hat p_1+m_1)
\hat\Pi_{S=1,0}(\hat p_2-m_2)\times
\end{displaymath}
\begin{displaymath}
\gamma^\sigma(\hat r_1-p_2+m_2)\gamma^\lambda(\hat q_2-m_2)\hat\Pi_{S=1,0}\Bigr]D^{\beta\omega}(t),
\end{displaymath}
where $m_1$, $m_2$ are the masses of electron and muon correspondingly, $k_1=k$, $k_2=t-k$ are four-momenta
of virtual photons in one loop, $r_1=k$, $r_2=t-r$ are four-momenta
of virtual photons in other loop. $D^{\beta\omega}(t)$ is the propagator of axial-vector meson.
After taking the trace in leading order in $\alpha$ and a number of simplifications, the amplitude numerator 
for hyperfine splitting can be represented as:
\begin{equation}
N^c_{AV}=\frac{1}{3}k^2 r^2 {\bf k}^2 {\bf r}^2,
\label{f4}
\end{equation}
where the index (c) denotes the contribution of the amplitude in Fig.~\ref{pic1}(c).
For the purpose of further integration over loop momenta, we pass to the Euclidean space:
\begin{equation}
k^2\to -k^2,~~~r^2\to - r^2,~~~k_0^2\to -k_0^2=-k^2\cos^2\psi_1,~~~r_0^2\to -r_0^2=-r^2\cos^2\psi_2.
\label{f5}
\end{equation}

As a result of all transformations, two integrals over k and r are factorized, and the contribution to the 
interaction operator in momentum space can be represented as follows:
\begin{equation}
\Delta V^c=-\frac{64}{9}\frac{\alpha^2(Z\alpha)^2}{{\bf t}^2+M_A^2}\int\frac{d^4 k}{\pi^2}A(t^2,k^2,k^2)
\frac{(2k^2+k_0^2)}{k^2(k^2-2m_1k_0)}
\int\frac{d^4 r}{\pi^2} A(t^2,r^2,r^2)\frac{(2r^2+r_0^2)}{r^2(r^2-2m_2r_0)}
\label{f6}
\end{equation}

To calculate each of the integrals, it is necessary to know the form of the transition form factor $A(t^2,k^2,k^2)$
of $1^{++}$ meson to two photons,
which is one of the main structural elements of the formula \eqref{f6}.
At present we have only few experimental data on it \cite{L3C,L3Ca,aihara}.
The L3 Collaboration studied the reaction $e^+e^- \to e^+e^-\gamma^\ast\gamma^\ast 
\to e^+e^-f_1(1285)\to e^+e^- \eta \pi^+\pi^-$ in \cite{L3C} and
measured the $f_1(1285)$ transition form factor for the case when one of the photons
is real and another one is virtual.
In \cite{L3Ca} the production of $f_1(1420)$ was investigated by the L3 Collaboration in the reaction
$\gamma^\ast\gamma^\ast\rightarrow K_S^0K^{\pm}\pi^\mp$.
By using  these data, we can parameterize the transition form factor for the case of two photons with equal 
virtualities as in our previous work \cite{apm1}:
\begin{equation}
A(M_{A}^2,k^2,k^2)=A(M_{A}^2,0,0)F^2_{AV}(k^2),~~~F_{AV}(k^2)=\frac{\Lambda_A^4}{(\Lambda_A^2-k^2)^2}.
\label{f7}
\end{equation}

The effects of off-shellness for exchange by massive $f_1$ mesons might be important. 
This effect was investigated in \cite{dorokhov1,dorokhov3}, and in \cite{ls} a simple parametrization
was proposed. The simplest way to take it into account is the introduction of the exponential suppression 
factor \cite{ls}:
\begin{equation}
\frac{A(t^2,0,0)}{A(M_{A}^2,0,0)}\approx e^{(t^2-M_{A}^2)/M_{A}^2},
\label{f8}
\end{equation}
which gives the factor $\sim e^{-1}$ for $t^2\approx 0$.
The values of the form factors in \eqref{f7} for the case of $f_1(1285)$ and $f_1(1420)$ can be fixed 
from L3 data \cite{apm1}:
\begin{eqnarray}
&&A_{f_1(1285)\gamma^\ast\gamma^\ast}\left(M_{f_1(1285)}^2,0,0\right)=(0.266 \pm 0.043)~\mathrm{GeV}^{-2}, \nonumber\\
&&A_{f_1(1260)\gamma^\ast\gamma^\ast}\left(M_{f_1(1260)}^2,0,0\right)=(0.160 \pm 0.120)~\mathrm{GeV}^{-2}, \nonumber\\
&&A_{f_1(1420)\gamma^\ast\gamma^\ast}\left(M_{f_1(1420)}^2,0,0\right)=(0.193 \pm0.041)~\mathrm{GeV}^{-2}.
\label{f9}
\end{eqnarray}

\begin{table}[h]
\caption{\label{tb1} Hadronic light-by-light contribution to muonium HFS.
The top line in each cell corresponds to a vertical exchange, and the bottom line corresponds to a horizontal exchange.
}
\bigskip
\begin{tabular}{|c|c|c|c|c|c|}   \hline
Meson &Mass   & $I^G(J^{PC})$  & $\Lambda$ & $A(M^2,0,0)$  & $\Delta E^{hfs}(1S)$ \\
   & in MeV   &    &   in MeV  &     &   in Hz    \\  \hline
$f_1$&1281.9  & $0^+(1^{++})$    & 1040& 0.266  &-0.00028        \\     
    &      &       &                  &     $GeV^{-2}$         &  -0.00053               \\    \hline
$a_1$& 1260 & $1^-(1^{++})$ &1040   &   0.160  & -0.00011         \\     
     &    &    &                  &      $GeV^{-2}$        &    -0.00020             \\    \hline
$f_1$&1426.3  & $0^+(1^{++})$   &926&  0.193  & -0.00007       \\    
    &     &    &                  &      $GeV^{-2}$        &   -0.00015              \\    \hline
$\sigma$  & 550&$0^+(0^{++})$    & 2000 & -0.596  &0        \\     
    &       &      &                  &     $GeV^{-1}$          &    0.02701             \\    \hline
$f_0$  &980 &$0^+(0^{++})$ &2000   &   -0.085  & 0          \\     
     &   &     &                  &       $GeV^{-1}$        &    0.00023             \\    \hline
$a_0$  &980 &$1^-(0^{++})$   &2000&  -0.086  & 0      \\    
    &    &     &                  &       $GeV^{-1}$        &   0.00023              \\    \hline
$f_0$  &1370 &$0^+(0^{++})$   &2000&  -0.036  & 0       \\    
    &     &    &                  &     $GeV^{-1}$          &   0.00002              \\    \hline    
$\pi^0$&134.9768 & $1^-(0^{-+})$   &770&  0.025  & 0        \\    
    &   &      &                  &       $GeV^{-1}$         &    -0.00135             \\    \hline
$\eta$&547.862  & $0^+(0^{-+})$   &774&  0.024  & 0        \\    
    &   &      &                  &        $GeV^{-1}$         &        -0.00019         \\    \hline    
$\eta'$& 957.78 & $0^+(0^{-+})$   &859&  0.031  & 0       \\    
    &   &      &                  &       $GeV^{-1}$          & -0.00013                \\    \hline    
$f_2$& 1275.4 & $0^+(2^{++})$   &2000&  0.498  & 0        \\    
    &   &      &                  &              &       0.00006           \\    \hline     
Total contribution   & \multicolumn{4}{c}{0.0245~Hz}   &   \\            \hline
\end{tabular}
\end{table}

Using the dipole parameterization from \eqref{f7}
we can calculate sequentially analytically the integrals over all variables in the Euclidean space:
\begin{equation}
I_e=\int d^4k\frac{(2k^2+k_0^2)}{k^2(k^2-2k_0m_1)}\frac{\Lambda^8}{(k^2-\Lambda^2)^4}=
-\int\limits_0^\infty d k^2 L_e(k^2) A(0,k^2,k^2)=
\label{f10}
\end{equation}
\begin{displaymath}
-\frac{\pi^2\Lambda_A^2}{4(1-a_e^2)^{5/2}}\left[3\sqrt{1-a_e^2}-a_e^2(5-2a_e^2)\ln\frac{1+\sqrt{1-a_e^2}}{a_e}\right],
\end{displaymath}
\begin{equation}
L_e(k^2) =\frac{\pi^2}{8m_1^4}\left[ k^2 (k^2 - 6 m_1^2) - (k^2 - 8 m_1^2)
\sqrt{k^2(k^2 + 4 m_1^2)}\right],~~~a_e=\frac{2m_1}{\Lambda}.
\label{f11}
\end{equation}

The integral for the muon loop $I_\mu$ is obtained by replacing $m_1\to m_2$.
Thus, final contribution to the muonium HFS can be represented by the following analytical formula:
\begin{equation}
\Delta E^{hfs}_c(1S)=-\frac{64\alpha^2(Z\alpha)^5\mu^3 A(0,0,0)^2}{9\pi M_A^2\left(1+\frac{2W}{M_A}\right)^2}I_eI_\mu.
\label{f12}
\end{equation}

For numerical estimates of this contribution, we take three axial vector mesons with masses 1285 MeV, 1260 MeV and
1420 MeV.
Total numerical value of the contribution is presented in Table~\ref{tb1}.
We write out in Table~\ref{tb1} numerical values of the individual contributions to the nearest five digits 
after the decimal point, bearing in mind that smallest contributions are of this order.
The contributions to the hyperfine splitting of the ground state in muonium are expressed in 
Table~\ref{tb1} in Hz, meaning the formula for the relationship between energy and frequency of the form
$\Delta\nu^{hfs}=\Delta E^{hfs}/h$.

Let us further consider horizontal exchanges with axial vector mesons shown in Fig.\ref{pic1}(a,b).
In this case, the use of projection operators \eqref{f2} also makes it possible to reduce the product 
of various factors in the numerator to a common trace, which can be calculated for the sum of the amplitudes
in Fig.\ref{pic1}(a,b) as
\begin{equation}
N^{(a+b)}_{AV}=(k_1^2k_2^4+k_2^2k_1^4)(2\cos\Omega+\cos\psi_1\cos\psi_2)-k_1^3k_2^3(1+3\cos^2\Omega+
\cos^2\psi_1+\cos^2\psi_2),
\label{f13}
\end{equation}
\begin{equation}
\cos\Omega=\cos\psi_1\cos\psi_2+\sin\psi_1\sin\psi_2\cos\theta.
\label{f14}
\end{equation}

To immediately take the sum of the amplitudes in Fig.\ref{pic1}(a,b), we multiply the direct amplitude by the 
factor $(k_2^2+2k_2^0m_2)$, and the cross amplitude by the factor $(k_2^2-2k_2^0m_2)$. 
In addition, we have passed to the Euclidean space of variables $k_1$ and $k_2$.
After all transformations, the contribution of horizontal exchanges to the HFS of the spectrum will be determined by the following integral expression:
\begin{equation}
\Delta E^{hfs}_{AV,(a+b)}=\frac{16\alpha^2(Z\alpha)^5\mu^3\Lambda^2}{3\pi}\int_0^\infty dk_1\int\frac{ d\Omega_1}{\pi^2}
\int_0^\infty dk_2\int\frac{ d\Omega_2}{\pi^2}\frac{A(M_A^2,k_1^2,k_2^2)}{(k_1^2+a_e^2\cos^2\psi_1)}\times
\label{f15}
\end{equation}
\begin{displaymath}
\frac{A(M_A^2,k_1^2,k_2^2)}{(k_2^2+a_\mu^2\cos^2\psi_2)}\frac{N^{(a+b)}_{AV}}
{(k_1^2+k_2^2+2k_1k_2\cos\Omega+\frac{M_A^2}{\Lambda^2})},
\end{displaymath}
where $d\Omega_1=2\pi \sin^2\psi_1\sin\theta d\theta d\psi_1$, $d\Omega_2=4\pi \sin^2\psi_2 d\psi_2$.
Further, the calculation of these integrals is carried out numerically, and the results are presented 
in Table~\ref{tb1}.

\section{Contribution of scalar mesons}
\label{sec3}

Recent results on the properties of light scalar mesons \cite{pdg} show that they are being intensively studied, 
including decays into two photons. But the accuracy of measuring the decay width $\Gamma_{S\gamma\gamma}$ is currently not high.
Let us consider the contribution of scalar mesons to the interaction amplitudes and HFS, using 
the methods formulated in the previous section for constructing hadronic light-by-light scattering amplitudes.
The general parametrization of scalar meson $\rightarrow \gamma^\ast+\gamma^\ast$ 
vertex function takes the form \cite{pauk,zhou,borisuk,volkov}:
\begin{equation}
\label{f16}
T_S^{\mu\nu}(t,k_1,k_2)=4\pi\alpha\biggl\{
A(t^2,k_1^2,k_2^2)(g^{\mu\nu}(k_1\cdot k_2)-k_1^\nu k_2^\mu)+
\end{equation}
\begin{displaymath}
B(t^2,k_1^2,k_2^2)(k_2^\mu k_1^2 - k_1^\mu (k_1\cdot k_2))(k_1^\nu k_2^2-k_2^\nu
(k_1\cdot k_2))\biggr\},
\end{displaymath}
where $A(t^2,k_1^2,k_2^2)$, $B(t^2,k_1^2,k_2^2)$ are two scalar functions on three variables, 
$k_{1,2}$ are four momenta of virtual photons, t is the four-momentum of scalar meson.
The first term in \eqref{f16} represents transverse photons interaction, and the
second term represents longitudinal photons interaction.
In the leading order, the contribution of the structure function $A(t^2,k_1^2,k_2^2)$ is decisive.
$t$ is the four momentum of scalar meson which is equal to $(k_1+k_2)$ for the horizontal exchanges.
The numerator of the sum of the horizontal exchange amplitudes is equal to
\begin{equation}
N^{(a+b)}_{S}=k_1^2k_2^2\cos\Omega(\cos\Omega\cos\psi_1\cos\psi_2-1-\cos^2\psi_1-\cos^2\psi_2-\cos^2\Omega).
\label{f17}
\end{equation}
The total contribution of scalar mesons to the hyperfine structure is similar to expression \eqref{f15} 
and in euclidean space has the following integral form:
\begin{equation}
\Delta E^{hfs}_{S,(a+b)}=\frac{16\alpha^2(Z\alpha)^5\mu^3}{3\pi}\int_0^\infty dk_1\int\frac{ d\Omega_1}{\pi^2}
\int_0^\infty dk_2\int\frac{d\Omega_2}{\pi^2}\frac{A(M_S^2,k_1^2,k_2^2)}{(k_1^2+a_e^2\cos^2\psi_1)}\times
\label{f18}
\end{equation}
\begin{displaymath}
\frac{A(M_S^2,k_1^2,k_2^2)}{(k_2^2+a_\mu^2\cos^2\psi_2)}\frac{N^{(a+b)}_{S}}
{(k_1^2+k_2^2+2k_1k_2\cos\Omega+\frac{M_S^2}{\Lambda^2})},
\end{displaymath}
where for the parameterization of a function $A(M_S^2,k_1^2,k_2^2)$ for scalar meson 
we use the monopole form for variables $k_1^2$ and $k^2_2$) as in our work \cite{apm3}:
\begin{equation}
A(M_s^2,k_1^2,k_2^2)=A_S\frac{\Lambda^4}{(k_1^2-\Lambda^2)(k_2^2-\Lambda^2)}.
\label{f19}
\end{equation}
The $S\gamma\gamma$ coupling constant $A_S$ is related to the $S\to\gamma\gamma$ partial width \cite{apm3,volkov,vl}:
\begin{equation}
A_S=\sqrt{\frac{4\Gamma_{S\gamma\gamma}}{\pi\alpha^2 M_S^3}},
\label{f20}
\end{equation}
where $M_S$ is the mass of the scalar meson,
$\Gamma_{S\gamma\gamma}$ is the radiative width of the scalar meson.

The vertical exchange amplitudes for scalar mesons are also constructed. The structure of the interaction 
vertices is such that the vertical exchanges are suppressed in comparison with the horizontal ones by the 
degree of momentum $|{\bf t}|\sim\mu\alpha$ and therefore give a contribution of a higher order in $\alpha$, 
which we omit below.

\section{Contribution of pseudoscalar mesons}
\label{sec4}

The transition vertex of two virtual photons into pseudoscalar meson is determined only by one
structure function.
The effective interaction vertex of the $\pi^0$ meson (or other pseudoscalar mesons $\eta$, $\eta'$) 
and virtual photons can be expressed in terms of the transition form factor 
$F_{\pi^0\gamma^\ast\gamma ^\ast}(k_1^2,k_2^2)$ in the form:
\begin{equation}
\label{f21}
V^{\mu\nu}(k_1,k_2)=i\varepsilon^{\mu\nu\alpha\beta}k_{1\alpha}k_{2\beta}
A(t^2,0,0)F_{\pi^0\gamma^\ast\gamma^\ast}(k_1^2,k_2^2),~~~A(M_P^2,0,0)=\frac{\alpha}{\pi F_P},
\end{equation}
where the pseudoscalar meson decay constants $F_P$ are $F_\pi=0.0924$~GeV,  $F_\eta=0.0975$~GeV,
$F_{\eta'}=0.0744$~GeV. The pseudoscalar decay constants are related to the two photon partial width
$\Gamma(P\to\gamma\gamma)$ of the resonance by the equation:
\begin{equation}
\label{f22}
F^2_P=\frac{\alpha^2}{64\pi^3}\frac{M_P^3}{\Gamma(P\to\gamma\gamma)},
\end{equation}
$M_P$ is the mass of pseudoscalar meson. The precise measurement of decay width 
$\Gamma(\pi^0\to\gamma\gamma)=7.82\pm 0.14 (stat) \pm 0.17 (syst)$ eV
was carried out in \cite{prim}. The result $\Gamma(\eta\to\gamma\gamma)=520\pm 20 (stat) \pm 13 (syst)$ eV
was obtained in \cite{kloe2}.

The transition form factor $F_{\pi^0\gamma^\ast\gamma^\ast}(k_1^2,k_2^2)$ is normalized by the condition:
$F_{\pi^0\gamma^\ast\gamma^\ast}(0,0)=1$.
Typically, it uses a monopole-type parameterization based on the squared momentum of each 
virtual photon, inspired by the vector dominance model \cite{kloe2,cleo,babar,belle,persson}:
\begin{equation}
F_{\pi^0\gamma^\ast\gamma^\ast}(k_1^2,k_2^2)=\frac{\Lambda^4}{(k_1^2-\Lambda^2)(k_2^2-\Lambda^2)}.
\label{f19a}
\end{equation}

The form factors of the transition of pseudoscalar mesons into two photons have been studied experimentally 
by various collaborations \cite{cleo,babar,belle,persson}. Fitting the experimental data using function \eqref{f19a} 
gave the following values 
of the cutoff parameter: $\Lambda_\pi=0.770$ GeV, $\Lambda_\eta=0.774$ GeV, $\Lambda_{\eta'}=0.859$ GeV.
From the theoretical viewpoint $F_\eta$, $F_{\eta'}$ entering in \eqref{f22} should be considered as 
effective decay constants due to $\eta-\eta'$ mixing \cite{kroll}.

The general formula that determines the contribution to the ground state HFS from the horizontal exchange amplitudes 
can be represented in integral form in Euclidean space:
\begin{equation}
\label{f23}
\Delta E^{hfs}_{PS}=-\frac{\alpha^2(Z\alpha)^5\mu^3}{3\pi F_P^2}\int\frac{d^4 k_1}{k_1\pi^4}
\int\frac{d^4 k_2}{k_2\pi^4}[F_{\pi^0\gamma^\ast\gamma^\ast}(k_1^2,k_2^2)]^2\times
\end{equation}
\begin{displaymath}
\frac{N^{(a+b)}_{PS}}{(k_1^2+a_e^2\cos\psi_1^2)(k_2^2+a_\mu^2\cos\psi_2^2)
((k_1+k_2)^2+\frac{M_P^2}{\Lambda^2})},
\end{displaymath}
where the function in the numerator
\begin{equation}
\label{f24}
N^{(a+b)}_{PS}=\left(\cos\Omega+\cos^2\Omega \cos\psi_1\cos\psi_2-\cos^3\Omega+\cos\psi_1\cos\psi_2-
\cos\Omega\cos^2\psi_1-\cos\Omega\cos^2\psi_2\right)
\end{equation}
is obtained by calculating the trace, summing over the Lorentz indices in an expression like
\begin{equation}
\label{f25}
\varepsilon^{\mu\nu\alpha\beta}k_{1\alpha}k_{2\beta}\varepsilon^{\lambda\sigma\rho\omega}k_{1\rho}k_{2\omega}
Tr[\gamma^\lambda (p_1-k_1+m_1)\gamma^\mu \hat\Pi\gamma^\nu (-p_2+k_2+m_2)\gamma^\sigma\hat\Pi^+].
\end{equation} 
The index $(a + b)$ denotes the contribution of the diagrams $(a)$ and $(b)$ in Fig.~\ref{pic1}.

Subsequently, integrals in \eqref{f23} are calculated numerically, as in the case of scalar mesons with horizontal exchanges.

Turning to the vertical exchange amplitudes, it should be noted that they contain additional powers 
of the momentum t. So, for example, the numerator of the amplitude in Fig.~\ref{pic1}(c) is equal to 
\begin{equation}
\label{f25a}
N^{c}_{PS}=t^2k^2r^2-k^2(rt)^2-r^2(kt)^2+(kr)(kt)(rt)-k_0r_0(kt)(rt).
\end{equation} 
As a result, it turns out that vertical interaction contribution to the hyperfine splitting is of order
$\alpha^2(Z\alpha)^7$. Therefore, this contribution can be neglected.

\section{Contribution of tensor mesons}
\label{sec5}

The lowest tensor resonance is the spin 2 $f_2(1270)$ dominating in $\gamma\gamma\to\pi^+\pi^-,\pi^0\pi^0$
production.
The $f_2$ parameters extracted are $M_{f_2} = 1275.4$ MeV, $\Gamma_{f_2} = 185.8$ MeV and 
$\Gamma_{f_2\gamma\gamma} / \Gamma_{f_2} = (1.42 \pm 0.24 \times 10^{-5})$.
For tensor mesons consisting from light quarks the experimental analysis of decay angular
distributions for $\gamma\gamma$ cross sections to $\pi^+\pi^-$, $\pi^0\pi^0$, $K^+ K^-$ have shown
that the $J=2$ mesons are produced mainly in a state with helicity $\Lambda=2$ \cite{p3}. We will assume further 
that hadronic light-by-light scattering amplitude for tensor mesons is dominated be helicity $\Lambda=2$ exchange.
Then the amplitude of the process $\gamma^\ast+\gamma^\ast\to T$ (see Fig.~\ref{pic1}) can be parametrised as follows \cite{pauk}:
\begin{equation}
T^{T}_{\mu\nu\alpha\beta}(k_1,k_2)=4\pi\alpha \frac{k_1 k_2}{M_T} {\cal M}_{\mu\nu\alpha\beta}(k_1,k_2){\cal F}_{T\gamma^\ast \gamma^\ast}(k_1^2,k_2^2),
\label{f26}
\end{equation}
where ${\cal F}_{T\gamma^\ast \gamma^\ast}(k_1^2,k_2^2)$ is a transition form factor, $k_1$, $k_2$ are
four momenta of virtual photons,
\begin{equation}
\label{f27}
{\cal M}_{\mu\nu\alpha\beta}(k_1,k_2)=\biggl\{R_{\mu\alpha}(k_1,k_2)R_{\nu\beta}(k_1,k_2)+
\frac{1}{8(k_1+k_2)^2\left[(k_1k_2)^2-k_1^2k_2^2\right]}R_{\mu\nu}(k_1,k_2) \times
\end{equation}
\begin{displaymath}
\left[(k_1+k_2)^2(k_1-k_2)_\alpha-(k_1^2-k_2^2)(k_1+k_2)_\alpha\right]\times
\left[(k_1+k_2)^2(k_1-k_2)_\beta-(k_1^2-k_2^2)(k_1+k_2)_\beta\right]\biggr\},
\end{displaymath}
\begin{displaymath}
R_{\mu\nu}(k_1,k_2)=-g_{\mu\nu}+\frac{1}{X}\left[(k_1k_2)(k_1^\mu k_2^\nu+k_2^\mu k_1^nu)-
k_1^2 k_2^\mu k_2^\nu-k_2^2 k_1^\mu k_1^\nu\right],~X=(k_1 k_2)^2-k_1^2k_2^2.
\end{displaymath}

Then the electron-muon direct interaction amplitude via horizontal tensor meson exchange can be presented as
follows:
\begin{equation}
\label{f28}
i{\cal M}_{T}=\frac{\alpha^2(Z\alpha)^2}{16m_1^2m_2^2}\int\frac{d^4k_1}{\pi^2 (k_1^2)^2}\int\frac{d^4k_2}{\pi^2 (k_2^2)^2}
\frac{(k_1k_2)^2}{M_T^2}{\cal F}^2_{T\gamma^\ast \gamma^\ast}(k_1^2,k_2^2)
{\cal M}_{\mu\nu\alpha\beta}(k_1,k_2){\cal M}_{\sigma\lambda\rho\omega}(k_1,k_2)\times
\end{equation}
\begin{displaymath}
D_T^{\alpha\beta\rho\omega}(k_1+k_2)Tr\Bigl[\hat\Pi(\hat q_1+m_1)\gamma^\sigma S_e(p_1-k_1)\gamma^\mu
(\hat p_1+m_1)\hat\Pi(\hat p_2-m_2)\gamma^\nu S_\mu(-p_2+k_2)
\gamma^\lambda(\hat q_2-m_2)
\Bigr],
\end{displaymath}
where $S_e(p_1-k_1)$ and $S_\mu(-p_2+k_2)$ are the propagators of electron and muon.
The massive spin 2 propagator has the form:
\begin{equation}
D^{\mu\nu\alpha\beta}_T(k)=\frac{f^{\mu\nu\alpha\beta}}{k^2-M_T^2+i0},
\label{f28a}
\end{equation}
\begin{equation}
f^{\mu\nu\alpha\beta}=\frac{1}{2}\left(g^{\mu\alpha}g^{\nu\beta}+g^{\mu\beta}g^{nu\alpha}-g^{\mu\nu}g^{\alpha\beta}\right)+
\frac{1}{2}\left(g^{\mu\alpha}\frac{k^\nu k^\beta}{M_T^2}+g^{\nu\beta}\frac{k^\mu k^\alpha}{M_T^2}+
g^{\mu\beta}\frac{k^\nu k^\alpha}{M_T^2}+g^{\nu\alpha}\frac{k^\mu k^\beta}{M_T^2}\right)+
\label{f28b}
\end{equation}
\begin{displaymath}
+\frac{2}{3}\left(\frac{1}{2}g^{\mu\nu}+\frac{k^\mu k^\nu}{M_T^2}\right)\left(\frac{1}{2}g^{\alpha\beta}+
\frac{k^\alpha k^\beta}{M_T^2}\right).
\end{displaymath}

The crossed amplitude in Fig.~\ref{pic1}(b) has the similar structure.
After further simplifications of the numerator of the expression \eqref{f28} in the Form package, 
it takes the following form:
\begin{equation}
\label{f29}
N_T^{(a,b)}=k_1^0k_2^0-k_1k_2+\frac{1}{k_1^2k_2^2-(k_1k_2)^2}[k_1^2(k_2^0)^2(k_1k_2)+
k_2^2(k_1^0)^2(k_1k_2)-2k_1^2k_2^2k_1^0k_2^0].
\end{equation}
For the form factor of the transition of a tensor meson into two virtual photons, we use 
the monopole parametrization with respect to each square of the photon momentum of the form:
\begin{equation}
\label{f30}
{\cal F}_{T\gamma^\ast \gamma^\ast}(k_1^2,k_2^2)=\frac{A_{T\gamma^\ast \gamma^\ast}(M_T^2,0,0)\Lambda_T^4}
{(k_1^2-\Lambda_T^2)(k_2^2-\Lambda_T^2)},
\end{equation}
and the value of $A_{T\gamma^\ast \gamma^\ast}(M_T^2,0,0)$ is determined using the width $\Gamma_{T~\gamma\gamma}$
of the decay of the tensor meson into two photons:
\begin{equation}
\label{f31}
A_{T\gamma^\ast \gamma^\ast}(M_T^2,0,0)=\frac{2\sqrt{5\Gamma_{T~\gamma\gamma}}}{\alpha\sqrt{\pi M_T}}.
\end{equation}

Numerical value of the decay width of one tensor meson $f_2(1275)$ is taken from
\cite{pdg}.
After passing to the Euclidean space and a number of simplifications, we can represent total 
contribution in Fig.~\ref{pic1}(a+b) to the hyperfine structure of muonium in the integral form:
\begin{equation}
\label{f32}
\Delta E^{hfs}_T=\frac{128\pi \alpha^2(Z\alpha)^5\mu^3}{3M_T^2}\int_0^\infty k_1^2 dk_1
\int_0^\pi \frac{\sin^2\psi_1}{\pi^3}\int_0^\infty k_2^2 dk_2
\int_0^\pi \frac{\sin^2\psi_2}{\pi^3}\int_0^\pi \sin\theta d\theta
\end{equation}
\begin{displaymath}
\frac{A^2_{T\gamma^\ast \gamma^\ast}(M_T^2,0,0)N_T^{(a,b)}}{(k_1^2+1)^2(k_2^2+1)^2(k_1^2+a_e^2\cos^2\psi_1)(k_2^2+a_\mu^2\cos^2\psi_2)[(k_1+k_2)^2+\frac{M_T^2}{\Lambda^2}]}.
\end{displaymath}
The results of numerical calculation \eqref{f32} are presented in Table~\ref{tb1} only for one tensor meson
since the contribution of other mesons is negligible due to the small width $\Gamma_{T~\gamma\gamma}$.

\section{Conclusion}
\label{concl}

As is known, the last measurement of the hyperfine splitting of the ground state in muonium was 
carried out in 1999 \cite{lanl} with a record-breaking accuracy for those times up to hundredths of a kHz. 
In a recent paper \cite{museum}, the MuSEUM collaboration announced the start of new measurements of HFS in muonium 
and obtained a result that agrees with \cite{lanl}, but is still inferior to it in accuracy. It can be said that 
the planned increase in the accuracy of measuring HFS in muonium to 1 ppb \cite{jparc} opens a new stage in the theoretical 
study of this problem, which is connected with an increase in the accuracy of calculations of various corrections. 
It should be emphasized that theoretical work in this direction did not stop during the last two 
decades \cite{eides,eides1,sgk,eides2,eides1a}. 
Various high-order quantum electrodynamic contributions in $\alpha$ were calculated. A sharp increase 
in the experimental accuracy leads to the need to take into account in the theoretical calculations 
the contributions of other interactions, as is the case for the anomalous magnetic moment of the muon 
or the Lamb shift in muonic hydrogen. This work is devoted to the study of one of these new contributions, 
due to the production of hadrons in light-by-light scattering amplitudes.

Compared to previous work \cite{apm2002}, this study takes into account the contributions of light mesons 
of different spins both in horizontal-type diagrams (Fig.~\ref{pic1}(a,b)) 
and in amplitudes with vertical exchange  (Fig.~\ref{pic1}(c)) . The calculated 
contributions from various mesons are presented separately in Table~\ref{tb1}. 
For all mesons, the parameter $A(M^2,0,0)$, 
which is also presented in the Table~\ref{tb1}, plays an important role in numerical evaluation of the contribution. 
Numerical value of this parameter is related to the width of the meson decay into two photons, which 
is taken from various experiments. An analysis of available experimental data on the decay widths 
into two photons shows that the accuracy of their measurement is not high \cite{pdg}. Therefore, it is more correct 
to consider the results presented in Table~\ref{tb1} as possible estimates of contributions of this type.

For mesons for which the $\Gamma_{\gamma\gamma}$ value has not yet been fixed \cite{pdg}, 
the average values were taken from the available data.
But with pseudovector and pseudoscalar mesons, which make an important contribution to Table~\ref{tb1}, 
the situation with fixing $A(M^2,0,0)$
is more or less certain, so that the error of their obtained contributions does not exceed 30 percent.
Nevertheless, there is a significant scatter in experimental data for the width of the $\sigma$ meson 
$\Gamma_{\sigma\gamma\gamma}$. 
In our calculations, we use for it a value of 4.5 keV. Since in the end it turns out that the contribution 
of this meson is the main one, we estimate total error of the calculation in Table~\ref{tb1} at 50 percent.
It should be noted also that the contribution of scalar meson depends on the type of form factor. 
We use for it a monopole parameterization \eqref{f19} based on the squared momentum of each photon, 
as in previous works \cite{apm3}. 
This parameterization is consistent with calculations of the form factor for the transition 
of a scalar meson into two photons, carried out within the framework of the quark model \cite{apm3,apm4,volkov}.
If the monopole parameterization is replaced by a dipole parameterization, the contribution is approximately halved.

It should be noted that the obtained contributions of pseudoscalar mesons improve our results due to more 
accurate numerical integration. The calculation formula \eqref{f23} is transformed in comparison with \cite{apm2002} 
in such a way that the contribution of both direct and cross horizontal amplitudes is taken into account 
at once. Numerically, the contributions of $\pi$, $\eta$, $\eta'$ mesons are among the most significant. 
As regards the contribution 
of axial vector mesons, they contribute from both types of exchanges (horizontal and vertical). 
The difference between our results on vertical exchanges of pseudovector mesons and work \cite{arkasha} is, 
in our opinion, that we take into account an additional reducing factor \eqref{f8}, the square of which 
just leads to a decrease in our contribution compared to \cite{arkasha} by an order of magnitude.
Contributions from exchanges of scalar and tensor mesons were not previously considered in \cite{apm2002,arkasha}.

As in the case of scalar mesons, there is a dependence of the results of calculating contributions 
on the type of transition form factor for both pseudovector and pseudoscalar mesons. From an experimental 
point of view, the best situation is with the form factor of the transition of a pseudoscalar meson into 
two photons \cite{cleo,babar,belle,persson}. As shown in \cite{persson}, the used by us parameterization \eqref{f19} 
is in good agreement with experimental data. 
The resulting error in calculating the contribution \eqref{f23} can be estimated at 10-15 percent.
There exist also data on the $Q^2$ dependence of transition form factor for $f_1\to\gamma\gamma^\ast$ \cite{L3C,L3Ca}.
In the analysis of the L3 data \cite{L3C,L3Ca} the single virtual transition form factor of the axial
vector mesons has been modelled by a dipole ansatz. 
In the case of two virtual photons, we use a form factor model in the form of a product of two such 
dipole functions. The error in calculating the contribution of pseudovector mesons can be estimated 
at 30 percent using this form of representation of the form factor and parameters \eqref{f9}.

Total contribution of all mesons to the hyperfine splitting turned out to be positive. Although the 
contributions of axial vector and pseudoscalar mesons are negative, there is a positive contribution 
of the $\sigma$ meson, which exceeds all previous ones in magnitude. The resulting value of 0.025 Hz can be 
regarded as an estimate of this small hadronic effect.

\begin{acknowledgements}
The authors are grateful to A.E. Radzhabov and A.S. Zhevlakov for useful discussions.
This work is supported by Russian Science Foundation (grant No. RSF 23-22-00143).
\end{acknowledgements}

\end{document}